# Direct CO$_2$ Electroreduction from Carbonate

*Yuguang C. Li[1†], Geonhui Lee[1†], Tiange Yuan[2], Ying Wang[1], Dae-Hyun Nam[1], Ziyun Wang[1], F. Pelayo García de Arquer[1], Yanwei Lum[1], Cao-Thang Dinh[1], Oleksandr Voznyy[2] and Edward H. Sargent[1*]*

[1]Department of Electrical and Computer Engineering, University of Toronto, 35 St George Street, Toronto, Ontario, M5S 1A4, Canada

[2]Department of Physical & Environmental Sciences, University of Toronto, Scarborough, 1065 Military Trail, Toronto, Ontario, M1C 1A4, Canada

**Corresponding Author**

* ted.sargent@utoronto.ca

The process of $CO_2$ valorization – all the way from capture/concentration of $CO_2$ to its electrochemical upgrade - requires significant inputs in each of the capture, upgrade, and separation steps. The gas-phase $CO_2$ feed following the capture-and-release stage and into the $CO_2$ electroreduction stage produce a large waste of $CO_2$ (between 80 and 95% of $CO_2$ is wasted due to carbonate formation or electrolyte crossover) that adds cost and energy consumption to the $CO_2$ management aspect of the system. Here we report an electrolyzer that instead directly upgrades carbonate electrolyte from $CO_2$ capture solution to syngas, achieving 100% carbon utilization across the system. A bipolar membrane is used to produce proton *in situ*, under applied potential, which facilitates $CO_2$ releasing at the membrane:catalyst interface from the carbonate solution. Using an Ag catalyst, we generate pure syngas at a 3:1 $H_2$:CO ratio, with no $CO_2$ dilution at the gas outlet, at a current density of 150 mA/cm$^2$, and achieve a full cell energy efficiency of 35%. The direct carbonate cell was stable under a continuous 145 h of catalytic operation at ca. 180 mA/cm$^2$. The work demonstrates that coupling $CO_2$ electrolysis directly with a $CO_2$ capture system can accelerate the path towards viable $CO_2$ conversion technologies.

**TOC GRAPHICS**

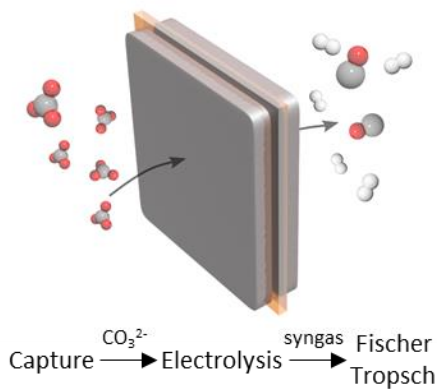

$CO_2$ capture systems often use alkali hydroxide solution to form alkali carbonate, and this requires additional energetic steps to dry and calcite the carbonate salt to generate a pure gas-phase $CO_2$ stream for the subsequent electrolysis reaction.[1-2] Direct electrochemical reduction of carbonate from the $CO_2$ capture solution could bypass the energy-intensive calcination step and significantly reduce the carbon footprint of the $CO_2$-to-products process.

This also addresses several limitations in the state-of-the-art $CO_2$RR systems: $CO_2$ waste arising due to the conversion of $CO_2$ gas into carbonate anions, especially in alkaline solutions.[3-4] Carbonate anions travel through an anion exchange membrane (AEM), along with some $CO_2$RR products, and be oxidized at the anode.[5] Additionally, as much as 80% of the input $CO_2$ gas may simply exit the electrolysis cell unreacted: many systems exhibit low single-pass utilizations even along the input-to-output gas channel.[6] As illustrated in Figure 1a, with the loss of $CO_2$ through carbonate formation, electrolyte crossover, and low single pass conversion efficiency, the utilization of carbon is low in many present-day $CO_2$RR electrolyzer designs.

We focused herein on carrying out $CO_2$RR electrolysis using carbonate solution directly as the carbon supply. We document 100% carbon utilization of input-carbon-to-products, evidenced by the lack of gaseous $CO_2$ at the reactor outlet. We do so by leveraging the facile acid/base reaction between proton and carbonate anion. We design an electrolysis system that generates $CO_2$ *in situ* from carbonate to initiate $CO_2$RR. Figure 1b shows the conventional/prior catalyst-membrane approach that uses a membrane-electrode-assembly (MEA) design. Here we instead use a bipolar membrane (BPM) which dissociates water to generate proton and hydroxide and directs them to the cathode and anode respectively.[7-9] Carbonate electrolyte circulates to the cathode via a peristaltic pump. Under applied potential conditions, the BPM proton reacts with carbonate to generate $CO_2$ near the membrane:cathode interface (Figure 1b and Video S1) which is reduced to value-added products via normal CO₂RR. The chemical balance of the full system is presented in Figure 1c.

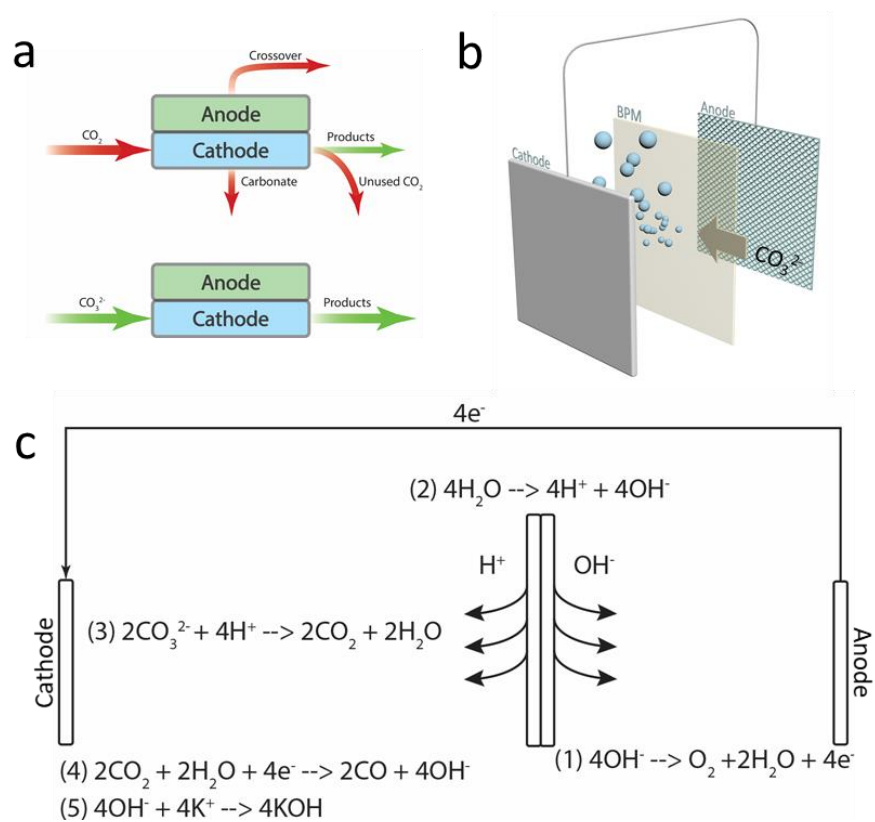

**Figure 1.** (a) Carbon loss mechanisms in a CO$_2$ electrolysis cell with gas-fed CO$_2$. (b) illustration of the bipolar membrane generating CO$_2$ *in situ* via the acid/base reaction of proton and carbonate ion. (c) Full chemical balance of the direct carbonate electrolysis cell with BPM.

We evaluated performance using Ag electrocatalysts (Figure S1) and Cu electrocatalysts (Figure S2) in 1 M K$_2$CO$_3$ electrolyte. The catholyte in Figure 2a-c was purged with N$_2$ to ensure that there is no dissolved CO$_2$. Ni foam was used as the anode with 1 M KOH electrolyte, a non-precious catalyst in an alkaline condition, favorable for the oxygen evolution reaction. All studies herein report the full cell voltage - which includes the series resistance, transport and kinetic overpotentials, from the cathode, anode and membrane – as seen for example in Figure 2a. The onset full-cell potentials for both Ag and Cu catalysts were observed at ca. 2.2 V, with Ag showing faster kinetics at higher applied potentials. For the Ag catalyst (Figure 2b), the CO Faradaic

Efficiency (FE) ranges from 28% to 12% at the applied current densities of 100 mA/cm$^2$ to 300 mA/cm$^2$, with the remainder of the FE being hydrogen. This yields a syngas ratio (H$_2$:CO) from ca. 2.5 to 7, suitable as feedstock to the Fischer-Tropsch (FT) reaction.[10] Since the source of carbon in this reaction is carbonate - a liquid phase reactant - the gas product exiting the electrolysis cell is pure syngas with small amount of moisture. Gas chromatography confirms no CO$_2$ is detected from the gas outlet stream. The full cell energy efficiency (EE) is 35% at 150 mA/cm$^2$, where we have included the contributions of both CO and H$_2$.

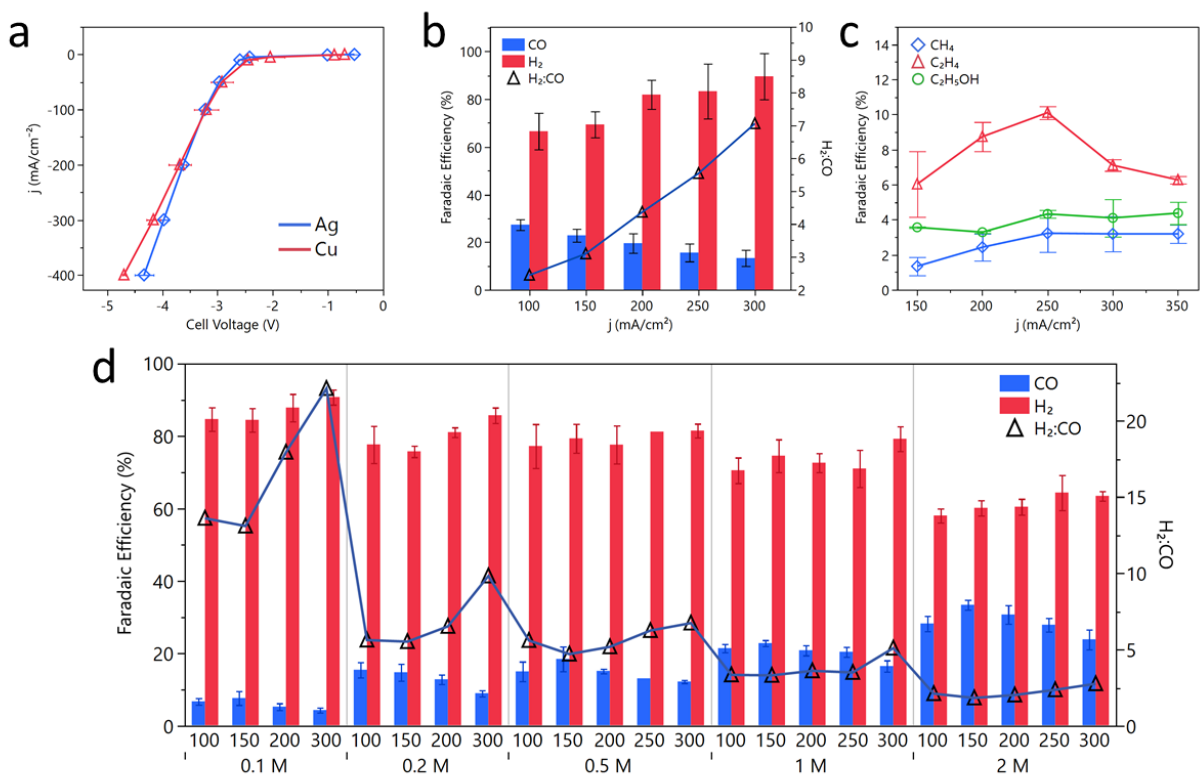

**Figure 2.** Performance of the direct carbonate electrolysis cell. (a) Full cell j-V curve with Ag and Cu catalyst. (b) Product distribution for the Ag catalyst. H$_2$ and CO are the major products, summing up to ~100% of the total FE. (c) Product distribution for the Cu catalyst. Propanol, formate and acetate are detected as well in a small amount. Figure (a) – (c) are conducted in 1 M K$_2$CO$_3$ catholyte with nitrogen purging as controls to demonstrate the concept of *in situ* CO$_2$ generations. 1 M KOH and Ni foam were used at the anode. (d) The product distribution of an Ag catalyst under different applied current density (1$^{st}$ x-axis, mA/cm$^2$) in different concentration of KOH electrolyte (2$^{nd}$ x-axis) purged with CO$_2$ prior to reaction, simulating the product of a CO$_2$ capture solution.

With a Cu catalyst, ca. 10% FE of ethylene is detected, as well as a small amount of ethanol and methane. In total, 17% $CO_2RR$ to hydrocarbon products was achieved. The full product distribution is available in Table S1.

The BPM also offers the benefit of mitigating product crossover as a result of the electro-osmotic drag of the proton emerging from the membrane, opposing the direction of products migration from cathode to anode.[5, 11] Anolytes from the Cu catalyst experiments were checked, and no liquid products were detected on the anode side. With this system design, the carbon loss mechanisms in a typical flow cell are overcome: $CO_2$ reaction with electrolyte to form carbonate; product crossover in the AEM system; and low single pass $CO_2$ utilization.

We examined the compatibility of the direct carbonate electrolysis cell in different $CO_2$ capture solutions directly. $CO_2$ gas was bubbled into 0.1 to 2 M of KOH solutions, simulating an industrial $CO_2$ capture process, and the $CO_2$ purged electrolyte was tested for carbonate electrolysis, showed in Figure 2d. The pH of the capture solution after $CO_2$ purging was approximately 11, which indicates that carbonate is the primary carbon species after $CO_2$ capture. With an Ag catalyst, the CO FE performance was observed to increase directly with respect to the concentration of the KOH electrolyte. This is likely due to the increase of the capture-generated $K_2CO_3$ concentration. The best performance of the KOH-$CO_2$ capture electrolyte shows a few percentage improvements compared to the pure $K_2CO_3$ electrolyte (Figure 2b). This is likely due to the small amount of bicarbonate salt present in the solution, generating small amount $CO_2$ via chemical equilibrium, and also a small amount of dissolved $CO_2$, both giving additional sources of reactant.[12-13]

In the full system chemical balance provided in Figure 1c, carbonate is consumed as the source of carbon in the cathodic reaction, and hydroxide is generated: this has the effect of regenerating the $CO_2$ capture solution. A capture-and-electrolysis system design can therefore operate continuously: the KOH capture solution removes $CO_2$ from the air or flue gas, forming carbonate; the carbonate electrolyte is then reduced to form value-added products via electrolysis with high carbon utilization; and the capture solution is thereby regenerated to restart the cycle.

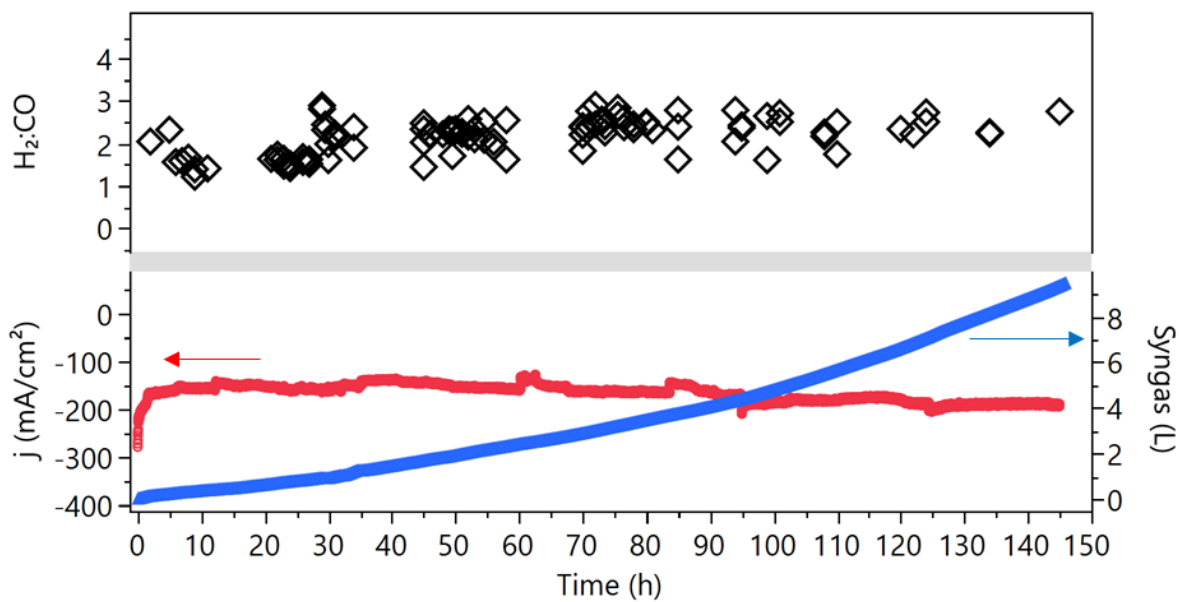

**Figure 3.** Stability evaluation of the direct carbonate electrolysis cell. $CO_2$ gas was first captured with KOH solution and transferred to an electrolysis bottle with no gas purging. The amount of gas produced from the electrolysis was measured with a mass flow meter and the ratio of $H_2$ and CO was monitored with GC injection. 1 M KOH and Ni foam were used at the anode. The cell was held at a constant potential of 3.8V.

We demonstrate a capture-electrolysis system in continuous operation for 145 hours with an Ag catalyst (in Figure 3). Two electrolyte bottles were used – one for capturing $CO_2$ gas directly with KOH electrolyte, and a second one for electrolysis. The carbonate capture solution and the electrolysis electrolyte are exchanged with a peristaltic pump (Figure S5). The electrolyte in the electrolysis bottle is pumped to the direct carbonate cell with no gas purging. Syngas generated from the reaction exits the bottle to a mass flow meter. The flow rate of gas products was recorded to calculate the total gas produced. During the 145 hours of electrolysis, the current density was stable at ca. 180 mA/cm² due to the pH balance and crossover prevention benefits offered by the BPM. The $H_2$:CO ratio also remains stable at between 2 and 3. Approximately 10 L of syngas were collected.

To assess the economics of the direct carbonate reduction, we calculated the energy cost per product molecule, considering the full process all the way from $CO_2$ capture and electrolysis to separation processes. We evaluated:

- alkaline flow cell[14]
- MEA cell with gas-fed $CO_2$
- direct carbonate cell explored herein.

Table 1 shows the summarizes the results (detailed calculations available in the SI). The total energy required to generate 1 mole of products is 4 times higher in the MEA cell with gas-fed $CO_2$ and 20 times higher for the alkaline flow cell. Figure 4 shows the energy capital per product molecule as a function of the $CO_2$ capture cost and the separation cost.[15-17] Even in the best-case scenario (low capture cost and low separation cost), the energy cost for $CO_2$RR in today's gas-fed $CO_2$ MEA cells is about two times higher than in the direct carbonate cell. Regeneration costs associated with removing carbonate from the electrolyte and from the anodic side add further to the expense of producing fuels and feedstocks in the gas-fed $CO_2$ MEA cell.

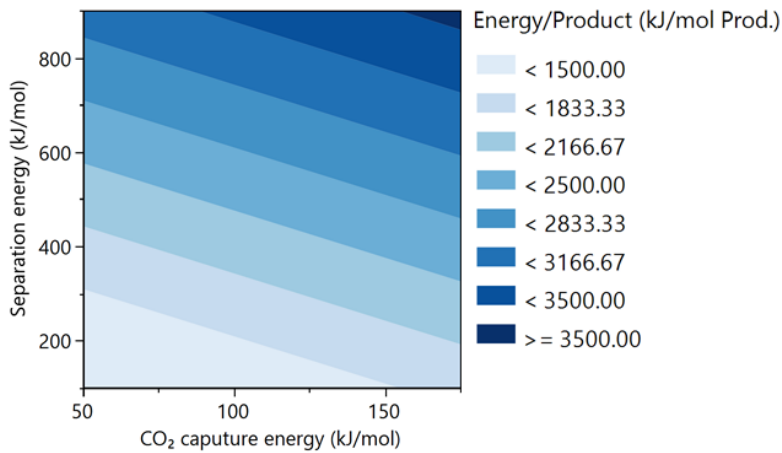

**Figure 4.** Technoeconomic analysis of the MEA cell with gas-fed $CO_2$ with different energy costs for $CO_2$ capture and different energy cost for products separation.

**Table 1.** The energy cost for the alkaline flow cell, $CO_2$ gas-fed MEA cell and direction carbonate cell. The cost of $CO_2$ capture was taken to be 178 kJ/mol[1] and the energy cost of separation is 500 kJ/mol.[15-16]

| Energy Capital | Flow Cell | MEA | Direct $CO_3^{2-}$ |
|---|---|---|---|
| **$CO_2$ Utilization** | 3 | 20 | 100 |
| carbonate formation (%) | 45 | 0 | 0 |
| crossover (%) | 2 | 30 | 0 |
| Exit $CO_2$ (%) | 50 | 50 | 0 |
| **$CO_2$ capture (kJ/mol of product)** | 5943 | 892 | 0 |
| $CO_2$ required (mol) | 33 | 5 | 1 |
| **$CO_2$ Electrolysis (kJ/mol of product)** | 476 | 733 | 733 |
| EE (%) | 54 | 35 | 35 |
| **Separation (kJ/mol)** | 8333 | 1250 | 0 |
| **Energy/Product (kJ/mol of product)** | 14753 | 2874 | 733 |

A number of topics require further study and progress in the direct carbonate cell. The thermodynamic onset potential for $CO_2$ reduction to syngas is approximately 1.34 V, and the experimental onset potential is ca. 2.2 V. The overpotential is large compared to a water electrolyzer, which obtains 1 A/cm$^2$ using less than 1 V of full cell overpotential.[18] The optimization of each cell components will be required to increase the full cell EE further and thereby lower the energy consumption for $CO_2$RR. While the gas products generated in the direct carbonate electrolysis cell do not contain $CO_2$, moisture is present in the exit stream, and requires separation before the syngas is utilized. There are also several competing reactions on the cathodic side. When a proton is generated from the BPM, it can be reduced directly on the cathode, leading to HER; when $CO_2$ is generated from carbonate, it can react with KOH, forming carbonate again, instead of being reduced in $CO_2$RR; and the proton from the BPM can also simply react with KOH in the electrolyte to form water. The penalties for these side reactions are reflected in less-than-100% total Faradaic efficiencies seen herein. The study of syngas in this report benefits from an industrially chosen preference of 30% $CO_2$-to-CO mixed with $H_2$, thus fits well with the finite FE to CO;[19] but future studies of carbonate-to-products will benefit from further insights, progress, and innovation to other higher value products in better conversion efficiency.

The system design herein achieves direct carbonate conversion via the acid/base reaction of proton and carbonate, which generates an *in-situ* source of $CO_2$, enabled by the use of a bipolar membrane. The device operated continuously for 145 hours and generated pure syngas in an optimal ratio suited for subsequent FT reaction. A faradaic efficiency of 17% of total carbonate-to-hydrocarbon products was also achieved with a Cu catalyst. This study demonstrates the direct implementation of carbonate to $CO_2$RR products from a $CO_2$ capture solution as input and a gas product suitable for the FT reaction as output. It enables direct $CO_2$ utilization from air or flue gas capture to hydrocarbon products.

ASSOCIATED CONTENT

**Supporting Information**. The Supporting Information is available free of charge on the ACS Publications web site at DOI:XXX

Experimental details and supplementary Figures S1-5

AUTHOR INFORMATION

**Corresponding Authors**

ted.sargent@utoronto.ca

**Author Contributions**

†These authors contributed equally to this work

**Notes**

The authors declare no competing financial interest.

ACKNOWLEDGMENT

The authors would like to acknowledge funding support from the Canadian Institute for Advanced Research (CIFAR), the Ontario Research fund and the Natural Sciences and Engineering Research Council (NSERC).